\documentclass[12pt]{iopart}

\usepackage{iopams}
\usepackage{graphicx}
\usepackage{bm}
\usepackage{cite}
\usepackage{color}

\begin{document}

\title[]{Gradiometric flux qubits with tunable gap}

\author{M J Schwarz$^1$, J Goetz$^{1}$, Z Jiang$^{1,2}$, T Niemczyk$^{1}$, F Deppe$^{1,2}$, A Marx$^{1}$ and R Gross$^{1,2}$}

\address{$^1$ Walther-Mei{\ss}ner-Institut, Bayerische Akademie der Wissenschaften, D-85748 Garching, Germany}
\address{$^2$ Physik-Department, Technische Universit\"{a}t M\"{u}nchen, D-85748 Garching, Germany}

\eads{\mailto{manuel.schwarz@wmi.badw.de}, \mailto{rudolf.gross@wmi.badw.de}}

\begin{abstract} %204 words
For gradiometric three-Josephson-junction flux qubits, we perform a systematic study on the tuning of the minimal transition frequency, the so-called qubit gap. By replacing one of the qubit's Josephson junctions by a dc SQUID, the critical current of this SQUID and, in turn, the qubit gap can be tuned \textit{in situ} by a control flux threading the SQUID loop. We present spectroscopic measurements demonstrating a well-defined controllability of the qubit gap between zero and more than 10\,GHz. This is important for tuning the qubit into and out of resonance with other superconducting quantum circuits in scalable architectures, while still operating it at its symmetry point with optimal dephasing properties. The experimental data agree very well with model calculations based on the full qubit Hamiltonian. From a numerical fit, we determine the Josephson coupling and the charging energies of the qubit junctions. The derived values agree well with those measured for other junctions fabricated on the same chip. We also demonstrate the biasing of gradiometric flux qubits near the symmetry point by trapping an odd number of flux quanta in the gradiometer loop. In this way, we study the effect of the significant kinetic inductance, thereby obtaining valuable information for the qubit design.
\end{abstract}

% 03.67.Lx Quantum computation architectures and implementations
% 74.50.+r Tunneling phenomena; Josephson effects
% 85.25.Cp Josephson devices

\vspace{2pc}
%\noindent{\it Keywords}: Article preparation, IOP journals
% Uncomment for Submitted to journal title message
%\submitto{\NJP}
\maketitle

\section{Introduction}
\label{section_introduction}

Superconducting quantum circuits are promising for the implementation of solid state quantum information systems \cite{Clarke:2008, Wendin:2007} and the realization of fascinating quantum-optical experiments in the microwave regime \cite{Bozyigit:2011, Eichler:2011, Menzel:2012}. In particular, the coupling of superconducting quantum two-level systems (qubits) to microwave resonators has been successful \cite{Wallraff:2004, Chiorescu:2004, Schoelkopf:2008}, resulting in the rapid development of the prospering field called circuit quantum electrodynamics (QED). In circuit QED, strong \cite{Wallraff:2004, Chiorescu:2004} and ultrastrong coupling \cite{Niemczyk:2010, Forn-Diaz:2010, Fedorov:2010} of superconducting quantum bits to the electromagnetic modes of high quality factor microwave resonators has been demonstrated. Circuit QED also has been used to generate non-classical states of light \cite{Hofheinz:2008, Fink:2008a}, to establish single artificial atom masing \cite{Astafiev:2007a}, to realize controlled symmetry breaking \cite{Deppe:2008,Niemczyk:2012}, or to implement quantum gates and algorithms \cite{Majer:2007, DiCarlo:2009}.

Nowadays, the most popular superconducting qubits are the cooper pair box \cite{Wallraff:2004, Nakamura:1999}, the transmon qubit \cite{Fink:2008a, Majer:2007, Koch:2007}, the phase qubit \cite{Hofheinz:2008, Martinis:1985, Devoret:2004}, the fluxonium \cite{Manucharyan:2009}, and the persistent current or flux qubit \cite{Chiorescu:2004, Mooij:1999, Chiorescu:2003, Orlando:1999}. For the implementation of circuit QED experiments, transmon and phase qubits have been used most often for several reasons. First, the relevant qubit parameters can be controlled within sufficiently narrow margins in the fabrication process. Second, a controlled coupling/decoupling to a microwave resonator acting as a quantum bus is possible by a fast change of the qubit's transition frequency. Third, the coherence properties of the qubit do not strongly degrade during such operations. Unfortunately, the original design of the flux qubit \cite{Mooij:1999} consisting of a superconducting loop intersected by three Josephson junctions cannot fulfill these requirements simultaneously. First, although the flux qubit's transition frequency $\omega_{\rm q}$ can be varied over a wide range by applying an external magnetic flux, the coherence time of the flux qubit rapidly decreases when tuning the qubit away from its symmetry point with minimum transition frequency $\omega_{\rm q} = \Delta$. Only at its symmetry point the flux qubit is well protected from the relevant $1/f$-noise and coherence times exceeding $1\,\mu$s can be reached \cite{Clarke:2008}. Second, for the flux qubit the minimal energy splitting $\hbar\Delta$ between the ground and excited state depends exponentially on the critical current $I_{\rm c}$ and capacitance $C_{\rm J}$ of the Josephson junctions \cite{Orlando:1999} and therefore is difficult to precisely control in fabrication. This does not allow to fabricate flux qubits with well-defined $\Delta$ values, which are, for example, close to the resonant angular frequency $\omega_{\rm r}$ of superconducting microwave resonators. On the other hand, flux qubits have specific advantages. First, the anharmonicity of flux qubits, that is, the separation of the excited state from the third level relative to $\omega_{\rm q}$, is by one or two orders of magnitude larger than for transmon and phase qubits, allowing for fast qubit operations without leakage to higher states. Second, flux qubits can be coupled ultrastrongly to resonators. Relative coupling strengths $g/\omega_{\rm r} > 0.1$, where $\omega_{\rm r}$ is the resonator frequency and $g$ the coupling strength, have been demonstrated \cite{Niemczyk:2010, Forn-Diaz:2010}. The reason is that the coupling is inductive for the flux qubit and capacitive for the transmon/phase qubit, respectively. As the inductive and capacitive coupling is proportional to $\Phi_0I_{\rm rms}$ and $2eV_{\rm rms}$, respectively, the ratio of the coupling strengths is $(h/4e^2) (I_{\rm rms}/V_{\rm rms}) = (h/4e^2) /Z_{\rm r} \sim 10$. Here, $\Phi_0 = h/2e$ is the flux quantum, $e$ the electron charge, $I_{\rm rms}$ and $V_{\rm rms}$ the zero-point current and voltage fluctuations in the resonator and $Z_{\rm r}$ the characteristic impedance of the resonator.

In order to overcome the drawback of fixed minimum energy splitting $\hbar\Delta$ in superconducting flux qubits, Orlando {\it et al.} \cite{Orlando:1999} proposed a modified flux qubit design, which subsequently has been implemented by Paauw {\it et al.} \cite{Paauw:2009} and meanwhile successfully used in several experiments, either in gradiometric \cite{Paauw:2009, Fedorov:2010, Fedorov:2011} or non-gradiometric design \cite{Zhu:2010, Poletto:2009, Gustavsson:2011, Zhu:2011a}. In this \emph{tunable-gap} flux qubit one of the Josephson junctions, the so-called $\alpha$-junction, is replaced by a small loop with two Josephson junctions. This dc superconducting quantum interference device (SQUID) acts as a junction whose critical current can be controlled by the flux threading the SQUID loop.  As a consequence, also the qubit gap $\Delta$ can be tuned in such a configuration. The additional control allows for a fast variation of the qubit transition frequency $\omega_{\rm q}$, while operating the flux qubit at its symmetry point where the coherence properties are optimum \cite{Fedorov:2010, Paauw:2009, Fedorov:2011}. In this article, we report on the fabrication and systematic study of tunable-gap gradiometric flux qubits, following the design proposed in \cite{Paauw:2009}. We emphasize that the combination of gradiometric design and tunable gap is especially suitable for the experimental realization of exciting proposals \cite{Porras:2012}. We show that the energy splitting at the symmetry point can be varied in a controlled way between values close to zero and $\Delta/2\pi > 10$\,GHz. In particular, our analysis extends to the case of multiple flux quanta trapped in the gradiometer loop. In this way, we obtain detailed insight into important design parameters and into the tuning mechanism. In section~\ref{section_flux_qubit}, we first introduce the foundations of fixed-gap and tunable-gap flux qubits required for the analysis of the experimental data. In particular, we discuss the possibility of flux-biasing gradiometric flux qubits at the symmetry point by freezing in an odd number of flux quanta during cool-down and the effect of the significant kinetic inductance of the narrow superconducting lines of the qubit loop. After briefly introducing the experimental techniques in section~\ref{section_techniques}, we present the experimental data and their analysis in section~\ref{section_results} before concluding with a brief summary in section~\ref{section_summary}.

\section{The flux qubit}
\label{section_flux_qubit}

In the following we briefly summarize the foundations of fixed-gap and tunable-gap persistent current or flux qubits as well as their gradiometric versions. We derive the relevant expressions used in the evaluation of our experimental data.

\subsection{The fixed-gap flux qubit}
\label{subsection_nontunable_flux_qubit}

\begin{figure}[tb]
\center{\includegraphics[width=0.99\columnwidth]{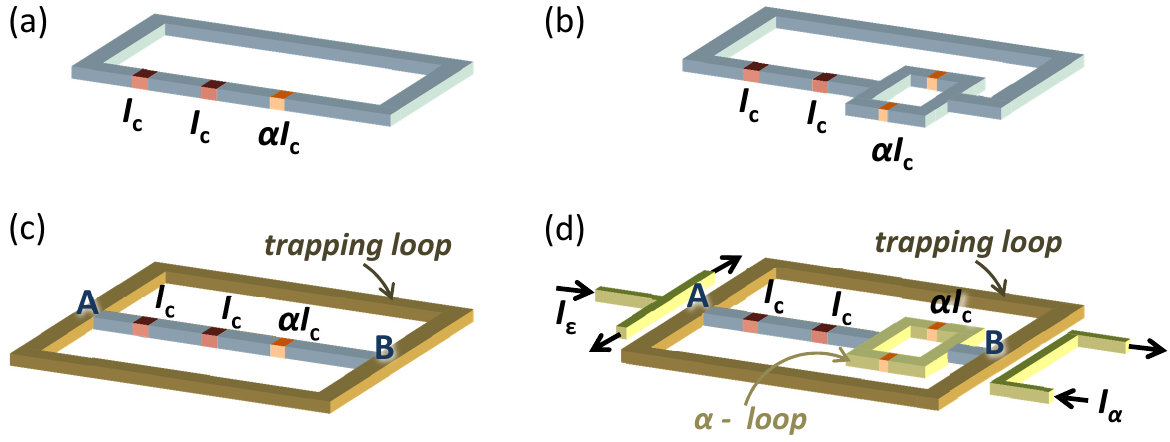}}
 \caption{
  Circuit schematics of (a) the three-Josephson-junction (3-JJ) flux qubit with $\alpha$-junction, (b) the simplest 3-JJ flux qubit with the tunable $\alpha$-junction realized by a dc SQUID, (c) the gradiometric 3-JJ flux qubit without tunable $\alpha$-junction, and (d) the gradiometric 3-JJ flux qubit with tunable $\alpha$-junction. The $\alpha$- and $\varepsilon$-lines can be used to change the magnetic frustration of the $\alpha$- and qubit loop independently.
 }
\label{Schwarz_NJP_Figure1}
\end{figure}

The simplest version of the flux qubit [cf. figure~\ref{Schwarz_NJP_Figure1}(a)] consists of a small superconducting loop with a diameter of the order of $10\,\mu$m intersected by three Josephson junctions (JJ) with lateral dimensions of the order of 100\,nm \cite{Mooij:1999}.  While two of these JJ have the same area (typically, $A_{\rm J}\simeq 0.03\,\mathrm{\mu m}^2$ in our experiments) and, hence, the same critical current (typically, $I_{\mathrm{c}}\simeq600$\,nA), the third JJ, the so-called $\alpha$-junction, has a reduced area $A_\alpha = \alpha A_{\rm J}$ with $\alpha \approx 0.6-0.8$, resulting in a reduced critical current $I_{{\rm c},\alpha}=\alpha I_{\rm c}$ and reduced junction capacitance $C_\alpha = \alpha C_{\rm J}$. Since $\alpha = \alpha_0$ is fixed in the fabrication process, the qubit gap $\Delta$ is also fixed. Consequently, this version of the flux qubit is called fixed-gap flux qubit. For $\alpha \approx 0.6-0.8$, the two-dimensional potential energy landscape of the flux qubit can be simplified. At the symmetry point, where the magnetic flux through the loop is equal to $(n+\frac{1}{2})\Phi_0$, with $n$ being an integer, the potential can be reduced to a one-dimensional double well \cite{Orlando:1999}. The two minima of this potential are associated with two degenerate persistent current states, corresponding to clockwise and counter-clockwise circulating persistent currents $\pm I_{\rm p}$. Due to the finite tunnel coupling of these states their degeneracy is lifted. The resulting symmetric and anti-symmetric superposition states form the ground and excited state of the flux qubit separated by the minimal energy splitting $\hbar\Delta$. Note that our terminology follows the most popular approach based on macroscopic quantum tunneling \cite{Martinis:1985}, however, interpretations based on the resistively and capacitively shunted junction model \cite{Marchese:2006, Marchese:2009, Blackburn:2012} have not been ruled out yet. In the basis of the persistent current states and near the symmetry point, the Hamiltonian describing the flux qubit can be written as  \cite{Orlando:1999}
\begin{equation}
	\label{Schwarz:qubit_Hamiltonian}
	 \mathcal{H} = \frac{1}{2} \hbar\varepsilon \sigma_z  -\frac{1}{2} \hbar\Delta \sigma_x \; .
\end{equation}
Here, $\sigma_z$ and $\sigma_x$ are the Pauli spin operators, $\hbar\varepsilon = 2I_{\rm p} \delta \Phi$ is the magnetic energy bias, and $\delta \Phi = \Phi_0 [f-(n+\frac{1}{2})]$ the deviation of the flux $\Phi$ threading the loop from a half-integer multiple of $\Phi_0$. The quantity $f=\Phi/\Phi_0$ is the magnetic frustration of the qubit loop and $n$ an integer. The transition frequency between ground and excited state can be written as 
\begin{equation}
	\label{Schwarz:qubit_transition_frequency}
	 \omega_{\rm q} = \sqrt{\varepsilon^{2}+\Delta^2} \; 
\end{equation}
and the qubit gap $\Delta$ becomes \cite{Orlando:1999}
\begin{equation}
\label{Schwarz:qubit_gap}
\Delta = \sqrt{\frac{4E_{\rm J} E_{\rm c} (4\alpha^2 -1)}{\alpha (1+2\alpha)}} \; \exp\left(- a(\alpha) \sqrt{4\alpha(1+2\alpha)\frac{E_{\rm J}}{E_{\rm c}}} \right)
 \; .
\end{equation}
Here, $E_{\rm J} = \hbar I_{\rm c}/2e$ is the Josephson coupling energy, $E_{\rm c} = e^2 /2C$ the charging energy and $a(\alpha) = \sqrt{1-(1/4\alpha^2)} - [\arccos (1/2\alpha)/2\alpha]$ with $a(\alpha) \simeq 0.15$ for $\alpha = 0.7$. Obviously, $\Delta$ is determined by the critical current $I_{\rm c}$ and the capacitance $C_{\rm J}$ of the Josephson junctions as well as by $\alpha = \alpha_0$. All these parameters are fixed by the fabrication process. Decreasing $\alpha$ from 1 to 0.5 results in a strong increase of the exponential factor. At the same time, the prefactor (attempt frequency) decreases from the plasma frequency of the JJ to zero, because the double well potential becomes a single well at $\alpha=0.5$. Since the exponential factor dominates within the major part of the interval $0.5<\alpha<1$, a strong increase of $\Delta$ is obtained by reducing $\alpha$.

It is evident from (\ref{Schwarz:qubit_Hamiltonian}) and (\ref{Schwarz:qubit_transition_frequency}) that it is possible to tune $\omega_{\rm q}$ by varying either $\varepsilon$ or $\Delta$. Varying the energy bias $\varepsilon$ is simply achieved by changing $\delta\Phi$ via the magnetic field generated by an external solenoid or a current fed through an on-chip control line. However, $\delta\Phi \ne 0$ causes a shift of the qubit operation point away from the anti-crossing point with minimal transition frequency $\Delta$. As the energy of the flux qubit is stationary with respect to small variations of the applied magnetic flux ($\partial \omega_{\rm q}/\partial\delta\Phi=0$) only for $f = \left(n+\frac{1}{2} \right)$, any shift away from this symmetry point makes the flux qubit more susceptible to magnetic flux noise and significantly deteriorates the coherence properties \cite{Deppe:2007, Kakuyanagi:2007}. Since the fast tuning of $\omega_{\rm q}$ of flux qubits is a prerequisite for numerous circuit QED experiments \cite{Mariantoni:2008, Leib:2012}, it is desirable to realize a tuning of $\omega_{\rm q}$ by a variation of $\Delta$.

\subsection{The tunable-gap flux qubit}
\label{subsection_tunable_flux_qubit}

According to (\ref{Schwarz:qubit_transition_frequency}), $\omega_{\rm q}$ cannot only be tuned by varying $\varepsilon$ but also by varying $\Delta$. This is advantageous, since the operation point of the qubit stays at the symmetry point with optimal coherence properties. Flux qubits with tunable $\Delta$ are called \emph{tunable-gap} flux qubits. As pointed out by Paauw {\it et al.} \cite{Paauw:2009}, an \emph{in situ} tunability of $\Delta$ is achieved by replacing the $\alpha$-junction by a small $\alpha$-loop containing two JJ [cf. figure~\ref{Schwarz_NJP_Figure1}(b)]. That is, the $\alpha$-junction is replaced by a dc SQUID. Then, the critical current $I_{{\rm c},\alpha}$ of the $\alpha$-loop and, in turn, the qubit gap $\Delta$ can be tuned by a control flux $\Phi_\alpha$ threading the $\alpha$-loop. If we choose the area of the junction in the $\alpha$-loop to $0.5\alpha_0 A_{\rm J}$, we obtain $I_{{\rm c},\alpha} = \alpha I_c$ with $\alpha= \alpha_0|\cos (\pi \Phi_\alpha/\Phi_0)|$. Successful implementations of this design have been reported recently \cite{Zhu:2010, Shimazu:2009a, Shimazu:2009b, Shimazu:2011a}.

The magnitude of $I_{\rm p}$ can be calculated to \cite{Orlando:1999}
\begin{equation}
\label{eq:Ip}
 I_{\rm p} = \pm I_{\rm c} \, \sqrt{1- \frac{1}{4\alpha^2}} \;
\end{equation}
yielding the $\alpha$-dependent transition frequency
 \begin{equation}
	\label{Schwarz:qubit_transition_frequency_alpha}
	 \omega_{\rm q} = \sqrt{4I_{\rm c}^2 \left[1-(1/4\alpha^2) \right] \delta\Phi^2/\hbar^2 +\Delta^2 (\alpha)} \;
\end{equation}
with $\Delta (\alpha)$ according to (\ref{Schwarz:qubit_gap}). We see that for $\alpha \rightarrow 0.5$ the persistent current $I_{\rm p}$ approaches zero.

\subsection{The gradiometric flux qubit}
\label{subsection_gradiometric_flux_qubit}

While the replacement of the $\alpha$-junction by an $\alpha$-loop allows for a tunable qubit gap $\Delta$, applying any flux to the $\alpha$-loop at the same time changes the flux threading the qubit loop and hence the energy bias $\varepsilon$ of the flux qubit. This is unintenional and has to be compensated. To keep the energy bias of the flux qubit constant during variations of $\Phi_\alpha$, a gradiometric design can be used. The gradiometric versions of a fixed-gap and tunable-gap flux qubit are shown in figures~\ref{Schwarz_NJP_Figure1}(c) and (d), respectively. 

For the tunable-gap gradiometric flux qubit of figure~\ref{Schwarz_NJP_Figure1}(d), an applied  homogeneous magnetic field changes $\Phi_\alpha$, but does not affect the energy bias of the flux qubit, since the screening currents in the two subloops of the eight-shaped gradiometric loop cancel each other on the central line. The immediate consequence is that an inhomogeneous magnetic field is required to adjust the energy bias $\varepsilon$ of the flux qubit. This inhomogeneous field can be generated by feeding a small current through the so-called $\varepsilon$-flux line, which couples asymmetrically to the qubit loop  [cf. figure~\ref{Schwarz_NJP_Figure1}(d)]. Furthermore, the outer ring of the gradiometric qubit, denoted as the trapping loop, can be used to trap an integer number of magnetic flux quanta, e.g. by cooling down below $T_{\rm c}$ in an applied magnetic field. This allows for a pre-biasing of the qubit near the symmetry point. We note, however, that the exact amount of flux threading the qubit loop and the $\alpha$-loop, respectively, depends on the ratio of the kinetic and geometric inductances. Since an understanding of this point is important for a controlled design of a gradiometric qubit with tunable gap, it is discussed in more detail in section~\ref{subsection_kinetic_inductance}. Within this work, we investigate both fixed-gap and tunable-gap gradiometric 3-JJ flux qubits. The former is an ideal model system to study the principle of flux biasing.

\subsection{Flux biasing}
\label{subsection_flux_biasing}

In this subsection we briefly address the flux biasing of gradiometric flux qubits by the trapping of magnetic flux in its outer loop, the so-called trapping loop [cf. figures~\ref{Schwarz_NJP_Figure1}(c) and (d)]. Flux biasing is based on the phase coherence of the superconducting state. The phase $\theta$ of the macroscopic wave function describing the superconducting state is allowed to change only by integer multiples of $2\pi$ along a closed integration path:
\begin{equation}
	\oint_{\Gamma} {\bm \nabla} \theta \cdot \rmd\bi{s} = 2\pi n  \; .
\label{eq:fluxquant1}
\end{equation}
In multiply connected superconductors this leads to the expression for the fluxoid quantization:
\begin{equation}
\oint_{\Gamma} \mu_0\lambda_{\rm L}^2 \bi{J}_{\rm s} \cdot \rmd\bi{s} +  \int_F \bi{B} \cdot \rmd\bi{F} = n\Phi_0  \; .
\label{eq:fluxoidquant1}
\end{equation}
Here, $\lambda_{\rm L}$ is the London penetration depth, $\mu_0$ the vacuum permeability, $\Gamma$ a closed integration path encircling the area $F$, $\bi{J}_{\rm s}$ the supercurrent density along $\Gamma$, and $\bi{B}$ the magnetic flux density. The second term on the left hand side represents the total magnetic flux $\Phi$ threading the area $F$. For superconductors with cross-sectional area large compared to the London penetration depth $\lambda_{\rm L}$, the first term vanishes, since one always can find an integration path deep inside the superconductor where the supercurrent density $J_{\rm s}=0$. This leads to the expression for flux quantization
\begin{equation}
 \frac{\Phi}{\Phi_0} \equiv f_{\rm tr} = n  \; ,
\label{eq:fluxquant2}
\end{equation}
saying that the total magnetic flux in a closed superconducting loop such as the trapping loop is quantized in units of $\Phi_0$.

The phase of the superconducting order parameter changes by $ 2\pi n$ around the closed trapping loop. Therefore, in the fully symmetric gradiometric qubit designs of figures~\ref{Schwarz_NJP_Figure1}(c) and (d) the trapping of an odd number $(2n+1)$ of flux quanta in the trapping loop leads to a phase difference of $(2n+1) \pi$ between the points A and B. This corresponds to a flux bias of $(n+\frac{1}{2}) \Phi_0$, i.e. a flux bias at the symmetry point. The biasing with trapped flux has the advantage that it is not affected by the finite noise of current sources required for the biasing with an external magnetic field. On the other hand, once a specific flux state has been frozen in, it can no longer be changed without heating the sample above $T_{\rm c}$. Therefore, in practice flux trapping is often used for pre-biasing at an operation point, while an additional magnetic field is used for making fast changes around this operation point. In order to enable such flux control the width of the superconducting line forming the trapping loop has to be made small enough (of the order of $\lambda_{\rm L}$) to allow for partial penetration of the applied magnetic field. In this case, the first term on the left hand side of (\ref{eq:fluxoidquant1}) becomes relevant. This term is related to the kinetic energy of the superconducting condensate or, equivalently, the kinetic inductance $L_{\rm k}$, whereas the second term is related to the field energy or, equivalently, the geometric inductance $L_{\rm g}$ of the trapping loop.

\subsection{The effect of the kinetic inductance}
\label{subsection_kinetic_inductance}

We next discuss the influence of the kinetic inductance $L_{\rm k}$, which is no longer negligible compared to the geometric inductance $L_{\rm g}$ of the trapping loop when the width of the superconducting lines is reduced to values of the order of $\lambda_{\rm L}$. In this case the first term on the left hand side of (\ref{eq:fluxoidquant1}) is no longer negligible. With the supercurrent $I_{\rm cir} = J_s S$ circulating in the trapping loop, we can rewrite this term as
\begin{equation}
 \oint_\Gamma \mu_0\lambda_{\rm L}^2 \bi{J}_s \cdot \rmd \bi{s} = \frac{\mu_0\lambda_{\rm L}^2}{S} \, I_{\rm cir} \ell = L_{\rm k}I_{\rm cir}    \; .
\label{eq:kin_inductance}
\end{equation}
Here, we have introduced the kinetic inductance $L_{\rm k} = \mu_0\lambda_{\rm L}^2 (\ell/S)$ of the trapping loop, with $\ell$ its circumference and $S$ its cross-sectional area. With $\Phi_{\rm k} = L_{\rm k} I_{\rm cir}$ and splitting up the total flux $\Phi$ into a part $\Phi_{\rm ex}$ due to an external applied field and a part $\Phi_{\rm g} = L_{\rm g} I_{\rm cir}$ caused by the circulating current in the trapping loop with geometric inductance $L_{\rm g}$, the fluxoid quantization condition (\ref{eq:fluxoidquant1}) reads as
\begin{equation}
 \frac{\Phi_{\rm k}}{\Phi_0} + \frac{\Phi_{\rm ex}}{\Phi_0} + \frac{\Phi_{\rm g}}{\Phi_0} \equiv f_{\rm k} + f_{\rm ex} + f_{\rm g} = n  \; .
\label{eq:fluxoidquant2}
\end{equation}
Introducing the parameter $\beta = L_{\rm g}/L_{\rm k}$, we obtain the expression for the net magnetic frustration of the trapping loop to
\begin{equation}
 f_{\rm tr, net} = f_{\rm ex} + f_{\rm g}  = \frac{1}{1+\beta}\, f_{\mathrm{ex}}+\frac{\beta}{1+\beta} \, n \; .
\label{eq:fluxoidquant3}
\end{equation}
The net magnetic frustration of the $\alpha$-loop in first approximation is obtained by multiplying with the area ratio $A_\alpha/A_\mathrm{tr}$ of the $\alpha$- and the trapping loop:
\begin{equation}
 f_{\alpha, \mathrm{net}} = \frac{A_\alpha}{A_\mathrm{tr}} \; f_{\mathrm{tr, net}} \; .
\label{eq:flux3}
\end{equation}
Here, we have neglected effects arising from the fact that the $\alpha$-loop is not centered in the trapping loop. If the geometric inductance is negligible ($\beta\ll1$), the contribution of the circulating screening current is negligible and $f_{\mathrm{tr, net}} \simeq f_{\mathrm{ex}}$. In this case, the superconducting lines cannot screen magnetic fields and we can change the magnetic frustration of the $\alpha$- and the trapping loop continuously by varying the applied magnetic field. In contrast, if the geometric inductance is dominant ($\beta\gg1$), the screening is so strong that we can no longer change the flux in the loop by varying the applied field. The frustration of the trapping loop is fixed to the value $f_{\mathrm{tr, net}}\simeq n$ frozen in during cool-down. This means that also the frustration of the $\alpha$-loop can no longer be changed continuously as desired.

With the net magnetic frustration (\ref{eq:flux3}), the critical current of the $\alpha$-loop is obtained to $I_{{\rm c},\alpha} (f_{\alpha, \mathrm{net}}) = \alpha (f_{\alpha, \mathrm{net}}) I_{\rm c}$ with
\begin{eqnarray}
 \alpha (f_{\alpha, \mathrm{net}}) & = & \alpha_{0} \left|\cos\left(\pi f_{\alpha\mathrm{,net}}\right)\right|
 \nonumber \\
 & = & \alpha_{0}\, \left|\cos\left(\pi\frac{A_{\mathrm{\alpha}}}{A_{\mathrm{tr}}} \; \left[ \frac{1}{1+\beta} \, f_{\rm ex} + \frac{\beta}{1+\beta} \, n \right] \right) \right| \; .
\label{eq:alphamax1}
\end{eqnarray}
For a suitable width of the superconducting lines and, hence, a suitable value of $\beta$, we can vary $\alpha$ both by changing $f_{\rm ex}$ via an external magnetic field and by changing the number $n$ of flux quanta frozen into the trapping loop during cool-down. For example, $n$ could be used for pre-biasing at a specific $\alpha$ value and the external magnetic field provided by a current sent through an on-chip control line for small variations around this value. The pre-biasing with trapped flux has the advantage that it is not affected by the noise added by the current source, while the variations with the on-chip control line can be very fast.

For zero applied magnetic field, (\ref{eq:alphamax1}) reduces to
\begin{equation}
 \alpha (f_{\alpha, \mathrm{net}})|_{f_{\rm ex} = 0} = \alpha_{0} \, \left|\cos\left(\pi\frac{A_{\mathrm{\alpha}}}{A_{\mathrm{tr}}} \; \frac{\beta}{1+\beta}\, n \right) \right| \; .
\label{eq:alphamax2}
\end{equation}
This expression applies to the experimental situation, where an odd number $(2n+1)$ of flux quanta is frozen into the trapping loop to bias the gradiometric flux qubit at its symmetry point and no additional external magnetic field is applied. Fixing $\alpha_0 \simeq 1$ by the fabrication process, we can change the number of trapped flux quanta to choose $\alpha$ in the desired regime $0.5<\alpha<1$.  Of course, flux trapping only allows for a step-wise variation of $\alpha$. For continuous and fast variations of $\alpha$, magnetic fields generated by external solenoids or on-chip control lines have to be used. For a typical value of $\beta \simeq 0.8$, we obtain $f_{\alpha, \mathrm{net}} \simeq 0.08\,n$ for $ f_{\mathrm{ex}} = 0$. This shows that we need only a small number of trapped flux quanta to significantly modify $\alpha$. Furthermore, we obtain $f_{\mathrm{tr, net}} = 0.55f_{\mathrm{ex}}$ for $n=0$, meaning that about half of the applied magnetic flux is shielded by the trapping loop.

\subsection{The gradiometer quality}
\label{subsection_gradiometer_quality}

A perfect gradiometer should be completely insensitive to a homogeneous magnetic field. However, in reality there are always imperfections such as slight differences of the areas $A_1$ and $A_2$ of the two subloops of the eight-shaped gradiometer and/or of the geometric inductances $L_{\rm g1}$ and $L_{\rm g2}$ and kinetic inductances $L_{\rm k1}$ and $L_{\rm k2}$ of the superconducting lines forming the subloops. Due to these imperfections there will be a finite imbalance $\delta f_{\rm imb}$ of the magnetic frustration of the two subloops. According to (\ref{eq:fluxoidquant2}), $\delta f_{\rm imb}$ can be expressed as
\begin{eqnarray}
 \delta f_{\rm imb} & = & \delta f_{\rm ex} + \delta f_{\rm g} + \delta f_{\rm k} = \frac{\delta \Phi_{\rm ex}}{\Phi_0} +  \frac{\delta\Phi_{\rm g}}{\Phi_0} + \frac{\delta\Phi_{\rm k}}{\Phi_0}
 \nonumber\\
 & = & \frac{\Phi_{\rm ex}}{\Phi_0} \frac{\delta A}{A} + \frac{I_{\rm cir}\delta L_{\rm g}}{\Phi_0} + \frac{I_{\rm cir}\delta L_{\rm k}}{\Phi_0}
 \; .
\label{eq:imbalance1}
\end{eqnarray}
With $I_{\rm cir} = (n-f_{\rm ex}) \Phi_0 /(L_{\rm g} + L_{\rm k})$ we can rewrite this expression to
\begin{equation}
 \delta f_{\rm imb} = f_{\rm ex} \underbrace{\left( \frac{\delta A}{A} - \frac{\delta L_{\rm g} +\delta L_{\rm k}}{L_{\rm g} + L_{\rm k}} \right)}_{\equiv 1/Q_{\rm grad,ex}} +\, n \underbrace{\left( \frac{\delta L_{\rm g} +\delta L_{\rm k}}{L_{\rm g} + L_{\rm k}} \right)}_{\equiv 1/Q_{\rm grad,n}}
 \; .
\label{eq:imbalance2}
\end{equation}
The total gradiometer quality $Q$ is given by $Q^{-1} = Q_{\rm grad,ex}^{-1} + Q_{\rm grad,n}^{-1}$. The first term describes imbalances of the frustration when a homogeneous external field is applied, and the second those when an integer number of flux quanta is frozen in. Obviously, the higher the $Q$ the lower is $\delta f_{\rm imb}$. As shown below, $Q$ values of the order of 500 are feasible.

We note that the $\varepsilon$-flux line shown in figure~\ref{Schwarz_NJP_Figure1}(c) generates different flux densities $B$ in the two subloops of area $A_{\rm tr,1} \simeq A_{\rm tr,2}$, leading to different amounts of total flux $\Phi_1 = \int_{A_{\rm tr,1}} B \rmd A$ and $\Phi_2=\int_{A_{\rm tr,2}} B \rmd A$. This results in the magnetic frustration
\begin{eqnarray}
 f_{12} & =  & \frac{\Phi_1 - \Phi_2}{\Phi_0}
 \;
\label{eq:imbalance3}
\end{eqnarray}
by the $\varepsilon$-flux line, which is used to change the energy bias $\varepsilon$ of the gradiometric flux qubit. Correspondingly, the deviation of $f_{12}$ from the value $(n+\frac{1}{2})$ at the symmetry point is
\begin{eqnarray}
 \delta f_{12} & =  & f_{12} - \left( n + \frac{1}{2} \right)
 \; .
\label{eq:imbalance4}
\end{eqnarray}

\section{Experimental techniques}
\label{section_techniques}

\begin{figure}[tb]
\center{\includegraphics[width=0.8\columnwidth]{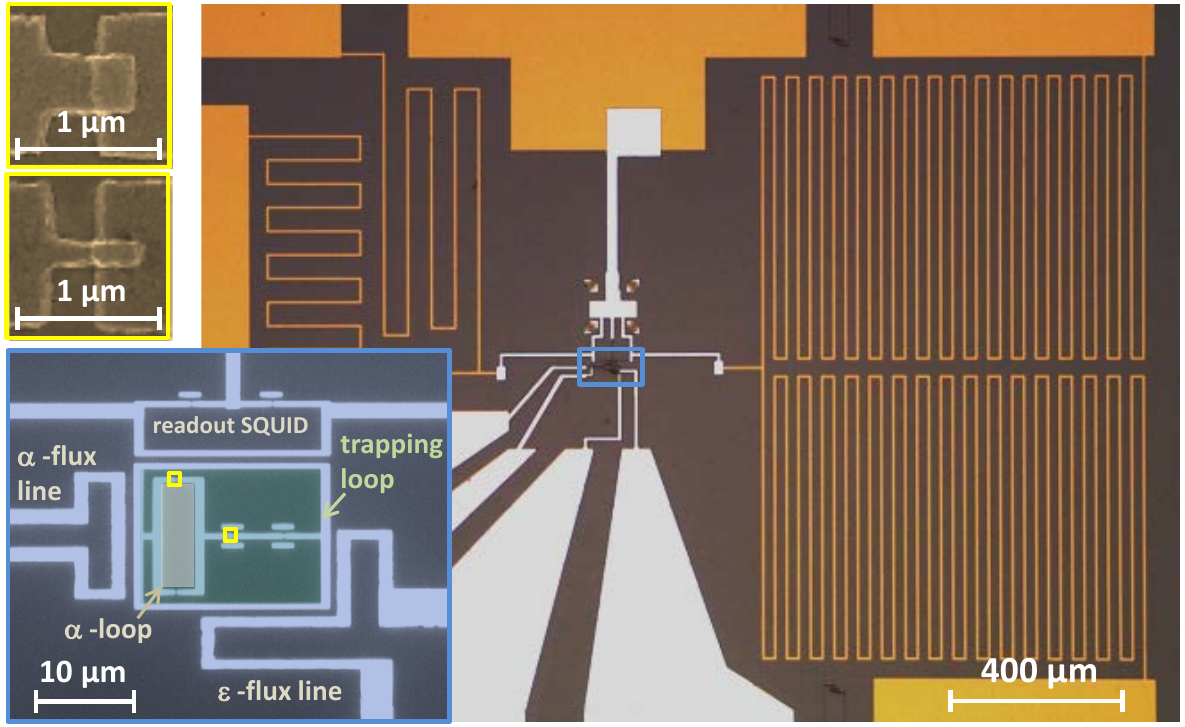}}
 \caption{
  Optical micrograph of the sample chip containing a tunable-gap gradiometric flux qubit with the biasing lines and filter structures. The large inset shows an enlarged view of the region marked with the blue rectangle: the gradiometric qubit with readout dc SQUID as well as the $\alpha$- and $\varepsilon$-flux lines used for tuning the frustration of the $\alpha$- and the qubit loop. The two small insets show scanning electron microscopy (SEM) images of a regular (top) and an $\alpha$-junction (bottom). The position of these junction are marked with yellow rectangles in the large inset.
 }
\label{Schwarz_NJP_Figure2}
\end{figure}

The flux qubits used in our study are based on Al thin film structures and Al/AlO$_x$/Al Josephson junctions fabricated by electron beam lithography and two-angle shadow evaporation on thermally oxidized silicon wafers. Details of the fabrication process can be found in \cite{Niemczyk:2009}. Figure~\ref{Schwarz_NJP_Figure2} shows optical and SEM micrographs of a sample chip with a tunable-gap gradiometric flux qubit. The qubit is surrounded by the readout dc SQUID and the control circuitry for the energy bias ($\varepsilon$-flux line) and the qubit gap $\Delta$ ($\alpha$-flux line). The insets show an enlarged view of the qubit region as well as SEM images of regular junctions and one of the junctions of the $\alpha$-loop, which has a reduced area $0.5\alpha_0 A_{\rm J}$ with $\alpha_0 \simeq 1$. The area of the regular junctions is $A_{\rm J} \simeq 0.03\,\mu$m$^2$.

All measurements have been performed in a dilution refrigerator with a base temperature of $30$\,mK. The qubit state is read out via a dc SQUID inductively coupled to the flux qubit \cite{Deppe:2007, Niemczyk:2009}. Qubit transitions between the ground and excited state can be induced by microwave radiation supplied via an off-chip antenna.  The continuous-wave (CW) microwave signal applied in our experiments is strong enough to saturate the qubit, leading to a 50\% population of ground and excited state. The trapping of flux quanta in the trapping loop of the gradiometric qubit is obtained by cooling down the circuit into the superconducting state in the presence of an appropriate magnetic field. Moreover, the sample can be heated up above $T_{\rm c}$ by applying a suitable heating current to an external heater located near the sample.

\section{Experimental results and discussion}
\label{section_results}

\subsection{Basic parameters}
\label{subsection_basic_parameters}

The critical current density of the Josephson junctions is determined to $J_{\rm c} (30\,{\rm mK}) = 1.5-3.5$\,kA/cm$^2$ by measuring the current-voltage characteristics (IVCs) of the readout SQUIDs fabricated on the same chip and by determining the junction area $A_{\rm J}$ by scanning electron microscopy. The $J_{\rm c}$ values can be varied by changing the oxidation process. For $J_{\rm c} = 2$\,kA/cm$^2$ and the typical junction area of $A_{\rm J}\simeq 0.03\,\mathrm{\mu m}^2$ we have $I_{\mathrm{c}}\simeq600$\,nA. The specific capacitance of the junctions is derived from the analysis of resonances in the IVCs of the readout SQUIDs \cite{Deppe:2004}. For junctions with $J_{\rm c} = 2$\,kA/cm$^2$, we find $C/A = (195 \pm 10)$\,fF/$\mu$m$^2$, resulting in $C_{\rm J}\simeq 6$\,fF for $A_{\rm J}\simeq 0.03\,\mathrm{\mu m}^2$. The geometric inductance $L_{\rm g}$ of the superconducting loops are estimated according to \cite{Terman:1950}. In order to estimate the kinetic inductance of the superconducting lines we use the dirty limit expression $L_{\mathrm{k}}=\hbar \rho_{\rm n} \ell/\pi \Delta_0 S$ \cite{Orlando:1991, Annunziata:2010}, where $\Delta_0=0.18$\,meV is the zero temperature energy gap of Al. The use of this expression is justified, since the mean free path in our 90\,nm thick Al films is limited by the film thickness and therefore is much smaller than the coherence length $\xi \simeq 1.5\,\mu$m of Al. The normal resistivity $\rho_{\rm n}$ is determined by suitable test structures fabricated on the same chip. For the cross-sectional area $S=500 \times 90$\,nm$^2$ of the superconducting line forming the trapping loop we obtain a kinetic inductance per unit length of $L_{\rm k}/\ell \simeq 1$\,pH/$\mu$m.

\subsection{The fixed-gap flux qubit}
\label{subsection_nontunable_flux_qubit_exp}

\begin{figure}[tb]
\center{\includegraphics[width=0.95\columnwidth]{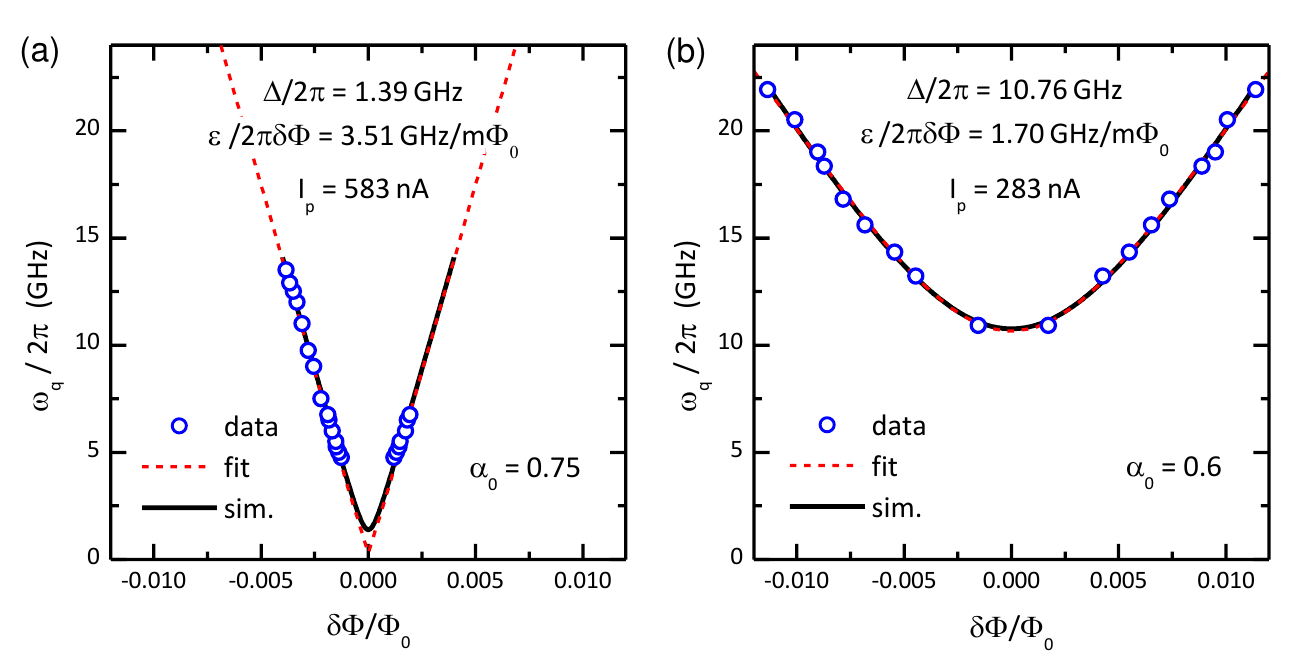}}
 \caption{
  Transition frequency $\omega_{\rm q}/2\pi$ plotted versus $\delta f = f-(n+\frac{1}{2})=\delta\Phi/\Phi_0 $ for two fixed-gap flux qubits with (a) $\alpha_0 = 0.75$ and (b) $\alpha_0 = 0.6$. Also shown is a two-parameter fit of the data (black lines) yielding $\Delta/2\pi$ and $I_{\rm p}$ and the result of a numerical simulation based on the diagonalization of the full qubit Hamiltonian. In (b) the result of the two-parameter fit and the simulation are almost indistinguishable.
 }
\label{Schwarz_NJP_Figure3}
\end{figure}

We first discuss the properties of fixed-gap, non-gradiometric flux qubits serving as reference samples. The qubit gap $\Delta$ and the persistent current $I_{\rm p}$ are determined by qubit spectroscopy \cite{Deppe:2007, Niemczyk:2009}. Figure~\ref{Schwarz_NJP_Figure3} shows typical spectra obtained for two 3-JJ flux qubits with fixed $\alpha$-junction by sweeping the qubit frustration $\delta f = f-(n+\frac{1}{2})=\delta \Phi/\Phi_0$ at fixed microwave frequency. The qubit state is read out repeatedly by the readout dc SQUID. Only at those $\delta f$ values where the microwave driving is resonant with the qubit transition frequency $\omega_{\rm q}$, a 50\% population of the excited state is detected. This manifests itself in characteristic peak and dip structures in the switching current $I_{\rm sw}$ of the readout SQUID at frequency dependent $\delta f$ values. Plotting these values versus the microwave frequency as shown in figure~\ref{Schwarz_NJP_Figure3} yields $\omega_{\rm q} (\delta f)$. Assuming that $J_{\rm c}$ has the same value for all three junctions, the value of $\alpha=\alpha_0=A_\alpha/A_{\rm J}$ can be determined from the measured area ratio. Then a two-parameter fit of (\ref{Schwarz:qubit_transition_frequency_alpha}) to the spectroscopy data yields $\Delta$ and $I_{\rm p}=\hbar\varepsilon/2\delta\Phi$. The spectra in figure~\ref{Schwarz_NJP_Figure3} are obtained for two flux qubits differing only in their $\alpha_0$ values. For $\alpha_0 = 0.75$ and 0.6, we obtain $\Delta/2\pi = 1.39$\,GHz and 10.76\,GHz and $I_{\rm p} = 583$\,nA and 283\,nA, respectively. Obviously, for $\alpha_0$ values closer to 0.5 (1.0) large (small) $\Delta$ and small (large) $I_{\rm p}$ values are obtained in agreement with (\ref{Schwarz:qubit_gap}) and (\ref{eq:Ip}). A consistency check can be made by calculating the $I_{\rm p}$ values from (\ref{eq:Ip}). Here, the unknown critical current $I_{\rm c} = J_{\rm c}A$ is estimated from the measured junction area and using the $J_{\rm c}$ value of the junctions of the readout SQUID. We obtain $I_{\rm p} = 619$\,nA and 310\,nA in good agreement with the values derived from the spectroscopy data.

We have also performed numerical simulations based on the diagonalization of the full qubit Hamiltonian using $E_{\rm J}$, $E_{\rm c}$ and $\alpha=A_\alpha/A_{\rm J}$ as input parameters. They are based on the $J_{\rm c}$ values derived from the IVCs of the readout SQUID and the measured junction areas. As shown in figure~\ref{Schwarz_NJP_Figure3}, there is very good agreement between the simulation result and the two-parameter fit for $\alpha =0.6$. However, significant deviations appear for $\alpha = 0.75$. The reason is that there are not enough data points around $\delta\Phi =0$, where the readout of the qubit state by the dc SQUID fails. This leads to large uncertainties in $\Delta$ for the two-parameter fit. Therefore, small $\Delta$ values tend to have larger error bars. Nevertheless, figure~\ref{Schwarz_NJP_Figure3} clearly demonstrates that the numerical simulation describes the experimental data very well.

\subsection{The fixed-gap gradiometric flux qubit}
\label{subsection_nontunable_gradiometric_flux_qubit_exp}

\begin{figure}[tb]
\center{\includegraphics[width=0.95\columnwidth]{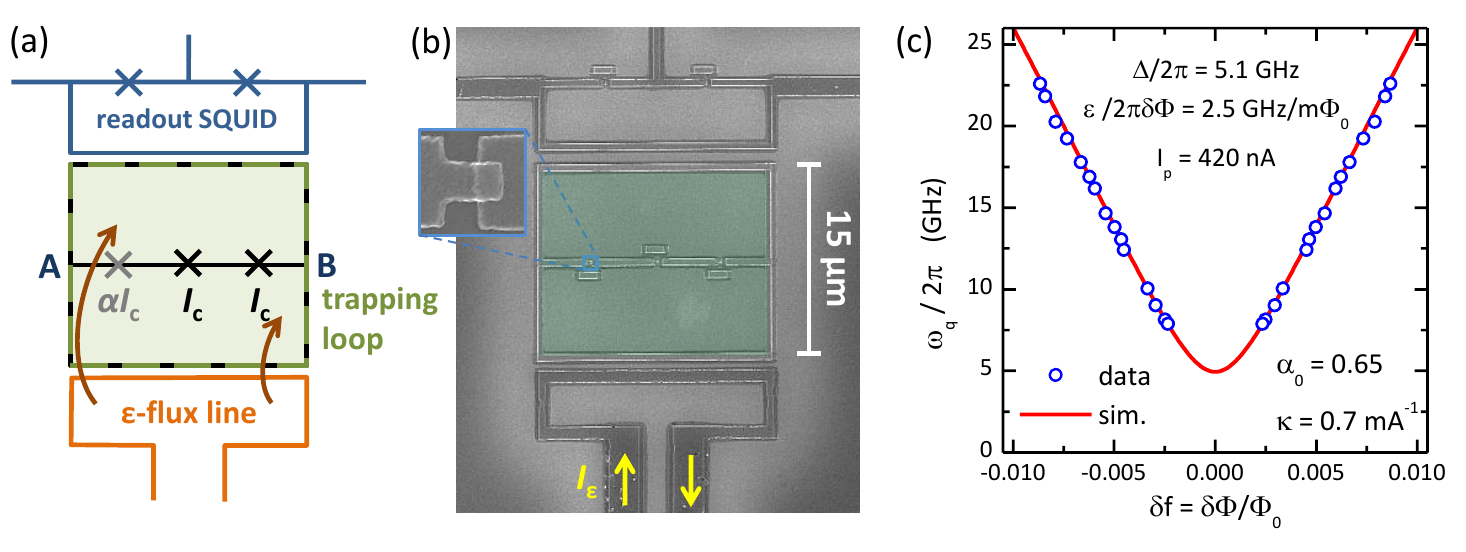}}
 \caption{
  (a) Circuit schematics of the fixed-gap gradiometric flux qubit with readout dc SQUID and $\varepsilon$-flux line. The outer loop of the flux qubit (broken olive line) forms the trapping loop. (b) Scanning electron microscope image of the implemented circuit. The inset shows an enlarged view of the $\alpha$-junction. (c) Transition frequency $\omega_{\rm q}/2\pi$ plotted versus $\delta f = f-(n+\frac{1}{2}) = \delta\Phi/\Phi_0$ for a fixed-gap gradiometric flux qubit with $\alpha_0 = 0.65$. Also shown is the result of a numerical simulation (red line) based on the diagonalization of the full qubit Hamiltonian with the listed parameters.
 }
\label{Schwarz_NJP_Figure4}
\end{figure}

We next discuss the properties of fixed-gap gradiometric flux qubits to demonstrate the operation of the gradiometric qubit design shown in figures~\ref{Schwarz_NJP_Figure4}(a) and (b). The flux qubit is biased close to its symmetry point by freezing in an odd number $(2n+1)$ of flux quanta in the trapping loop during cool-down. This results in a phase difference of $(2n+1) \pi$ between points A and B, equivalent to a flux bias of $(n+\frac{1}{2}) \Phi_0$ of the gradiometric flux qubit at its symmetry point. To change the energy bias $\varepsilon$ after cool-down, a spatially inhomogeneous magnetic field is required, which is generated by the current $I_\varepsilon$ sent through the $\varepsilon$-flux line. The qubit state is read out via the readout dc SQUID inductively coupled to the trapping loop of the qubit. The operation point of the readout dc SQUID can be optimized by applying a homogeneous magnetic field (e.g. by a solenoid) which does not affect the energy bias of the qubit due to its gradiometric design.

Figure~\ref{Schwarz_NJP_Figure4}(c) shows typical spectroscopy data of a gradiometric flux qubit with $\alpha_0=0.65$. Since we are measuring $\omega_{\rm q} (\delta I_\varepsilon)$ and not $\omega_{\rm q} (\delta f)$, the only problem in evaluating this data is to determine the calibration factor
\begin{equation}
 \kappa \equiv \frac{\partial \delta f}{\partial \delta I_\varepsilon} \; ,
\label{eq:transfer_function_k}
\end{equation}
where $\delta I_\varepsilon = I_\varepsilon - I_\varepsilon^{\rm sym}$ is the deviation of the current $I_\varepsilon$ sent through the $\varepsilon$-flux line from the value $I_\varepsilon^{\rm sym}$ needed for biasing the qubit at the symmetry point. This is done by calculating $\omega_{\rm q} (\delta f)$ by numerical simulations using $E_{\rm J}$, $E_{\rm c}$ and $\alpha=A_\alpha/A_{\rm J}$ as input parameters. The scaling factor $\kappa$ is then obtained by re-scaling the measured $\omega_{\rm q} (\delta I_\varepsilon)$ dependence to obtain optimum agreement with the simulation result. For the sample in figure~\ref{Schwarz_NJP_Figure4}(b), we obtain $\kappa=0.7$\,mA$^{-1}$, saying that a current of about 1\,mA results in $\delta f =1$. In general, the agreement between the experimental data and the simulation was found to be very good. The simulated values for the sample in figure~\ref{Schwarz_NJP_Figure4}(c) are $\Delta/2\pi = 5.1$\,GHz, $I_{\rm p} = 420$\,nA and $\alpha_0=0.65$. Again, we can make a consistency check by calculating the $I_{\rm p}$ value according to (\ref{eq:Ip}) as discussed above. We obtain $I_{\rm p} = 485$\,nA in good agreement with the value derived from the simulation.

We note that we can also trap an even number $2n$ of flux quanta in the trapping loop. In this case the phase difference between points A and B is $2 \pi n$. This corresponds to a flux bias of the gradiometric qubit by $2n \Phi_0/2 = n \Phi_0$ instead of $(n+\frac{1}{2}) \Phi_0$ for an odd number of trapped flux quanta. That is, the qubit is biased far away from its symmetry point and no qubit transitions should be observable. This is in full agreement with the experimental observation. We finally note that $I_{\rm cir} = (n-f_{\rm ex})\Phi_0/(L_{\rm g} + L_{\rm k})$ is not allowed to exceed the critical current of the trapping loop. For our samples, this fact  limits the number of trapped flux quanta to a maximum value of $10-15$.

\begin{figure}[tb]
\center{\includegraphics[width=0.95\columnwidth]{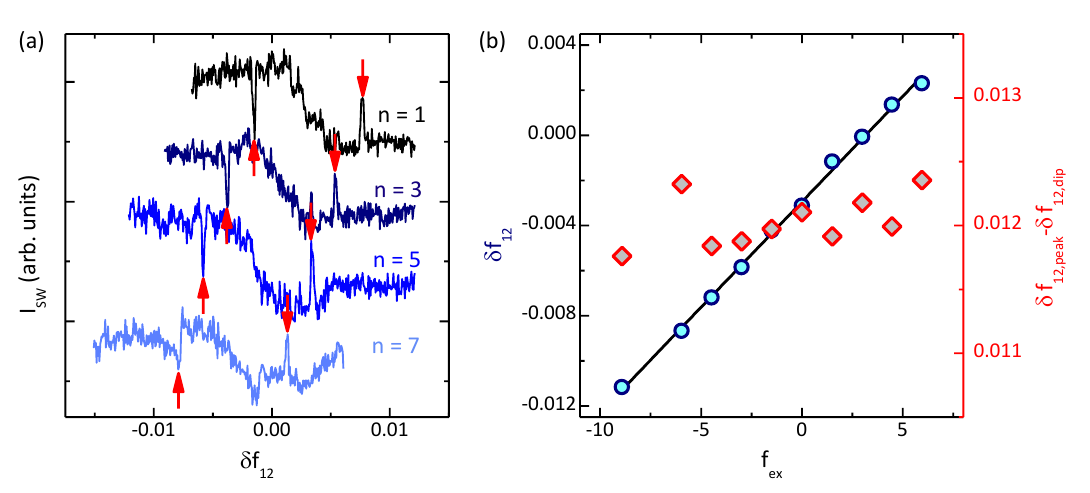}}
 \caption{
  (a) Switching current $I_{\rm sw}$ of the readout SQUID of a fixed-gap gradiometric flux qubit as a function of $\delta f_{12} =  f_{12} - \left(n+\frac{1}{2} \right)$ recorded for a fixed microwave frequency of 19.33\,GHz. The peak and dip positions mark those $\delta f$ values where $\omega_{\rm q}/2\pi = 19.33$\,GHz. (b) Frustration imbalance $\delta f_{12}$ as a function of the frustration $f_{\rm ex}$ generated by a homogeneous applied magnetic field. Also shown is the distance between the peak and dip positions in the $I_{\rm sw}(\delta f_{12})$ curves.
 }
\label{Schwarz_NJP_Figure5}
\end{figure}

We also use the simple fixed-gap gradiometric qubit to check the quality of the gradiometer discussed in section~\ref{subsection_gradiometer_quality}. Figure~\ref{Schwarz_NJP_Figure5}(a) shows the switching current of the readout SQUID as a function of $\delta f_{12} =  f_{12} - \left(n+\frac{1}{2} \right)$ [cf. (\ref{eq:imbalance4})] recorded for a fixed microwave frequency of 19.33\,GHz. The peaks and dips in the $I_{\rm sw}(\delta f_{12})$ curves mark the $\delta f_{12}$ positions where the qubit transition frequency $\omega_{\rm q}/2\pi = 19.33$\,GHz. On varying the number $n$ of trapped flux quanta, these positions shift due to the imperfect balance of the gradiometer. From the measured shift we derive $Q_{\rm grad,n} = 943 \pm 19$. In figure~\ref{Schwarz_NJP_Figure5}(b), $\delta f_{12}$ is plotted versus $f_{\rm ex}$ generated by a homogeneous applied magnetic field. From the measured slope the quality factor $Q_{\rm grad,ex} = 1076 \pm 16$ is determined. The total quality of the gradiometer is then $Q \simeq 500$, corresponding to a gradiometer imbalance of only 0.2\%. This means, that the qubit operation point is shifted by about 2\,m$\Phi_0$ when we apply a homogeneous field generating one $\Phi_0$ in the trapping loop. The measured quality factors are plausible. For example, the limited precision of the electron beam lithography process causes a finite precision $\delta A/A_{\rm tr}$ of the trapping loop area as well as $\delta S/S$ of the cross-sectional area and $\delta\ell/\ell$ of the length of the superconducting lines. The measured quality factor corresponds to $\delta A \simeq 0.2\,\mu$m$^2$, $\delta S \simeq 50\,$nm$^2$ or $\delta\ell \simeq 60$\,nm. These values agree well with the values expected for the precision of the fabrication process.

In figure~\ref{Schwarz_NJP_Figure5}(b) we also plot the distance between the peak and dip positions in the $I_{\rm sw}(\delta f_{12})$ curves. This distance is about independent of $f_{\rm ex}$. This demonstrates that the qubit potential is not affected by the homogeneous background field. In total, our results show that the gradiometric flux qubits can be fabricated in a controlled way and work as expected. The fact that the qubit operation point is not affected by a homogeneous background field allows us to integrate these qubits into large scale circuits where several qubits have to be operated and read out simultaneously without affecting each other.

\subsection{The tunable-gap gradiometric flux qubit}
\label{subsection_tunable_gradiometric_flux_qubit_exp}

In this subsection we discuss the results obtained with tunable-gap gradiometric flux qubits as sketched in figure~\ref{Schwarz_NJP_Figure1}(d). Besides a step-wise variation of $\alpha$ by freezing in an odd number of flux quanta in the trapping loop we can make a continuous variation of $\alpha$ by an applied magnetic field generated either by the current $I_{\rm coil}$ fed through an external solenoid or the current $I_\alpha$ fed through the on-chip $\alpha$-flux line. We first discuss the experiments using a homogeneous magnetic field of a solenoid placed underneath the sample. The homogeneous magnetic field generates the frustrations $f_{\rm tr,net}$ and $f_{\alpha,{\rm net}}$ of the trapping and $\alpha$-loop, respectively, which are given by (\ref{eq:fluxoidquant3}) and (\ref{eq:flux3}).

\begin{figure}[tb]
\center{\includegraphics[width=0.95\columnwidth]{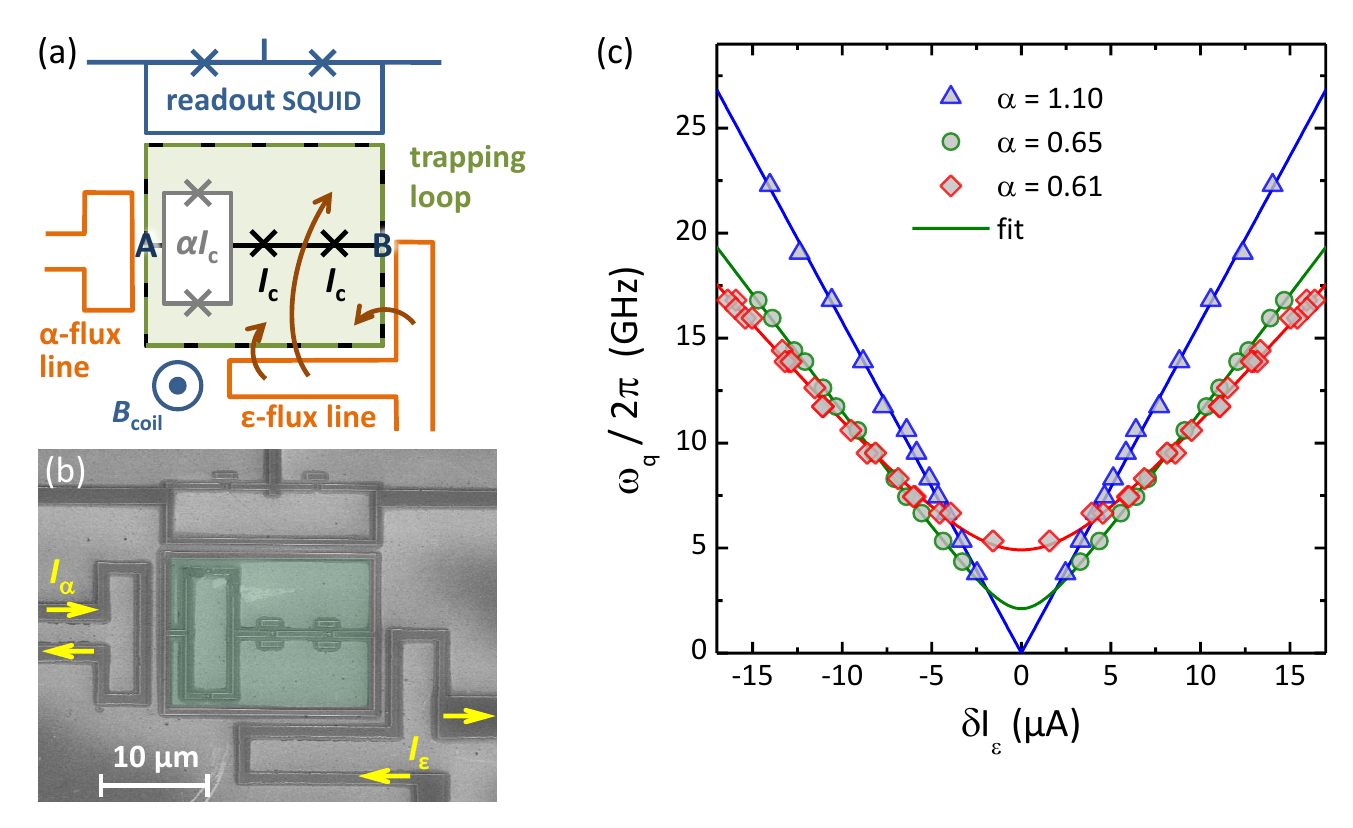}}
 \caption{
  (a) Circuit schematics of the tunable-gap gradiometric flux qubit with readout dc SQUID,  $\varepsilon$- and $\alpha$-flux line. The outer loop of the flux qubit (broken olive line) forms the trapping loop, the inner (grey line) the $\alpha$-loop. (b) Scanning electron microscopy (SEM) image of the implemented circuit. (c) Transition frequency $\omega_{\rm q}/2\pi$ plotted versus $\delta I_\varepsilon = I_\varepsilon - I_\varepsilon^{\rm sym}$ for three different $\alpha$ values for a tunable-gap gradiometric flux qubit with $\alpha_0 = 1.10$. Also shown is the result of a two-parameter fit.
 }
\label{Schwarz_NJP_Figure6}
\end{figure}

Spectroscopy data of a tunable-gap gradiometric flux qubit is shown in figure~\ref{Schwarz_NJP_Figure6}(c). The different $\alpha$ values were generated by the homogeneous magnetic field of the solenoid, whereas the flux trapped during cool-down was constant at a single flux quantum, i.e. $n=1$. We can fit the data by a two-parameter fit yielding $\Delta$ and the slope $\partial \omega_{\rm q}/\partial \delta I_\varepsilon$ at large $\omega_{\rm q}$ values. Here, $\delta I_\varepsilon = I_\varepsilon - I_\varepsilon^{\rm sym}$ is the deviation of the current $I_\varepsilon$ sent through the $\varepsilon$-flux line from the value $I_\varepsilon^{\rm sym}$ needed for biasing the qubit at the symmetry point. To derive the persistent current $I_{\rm p}= (\hbar/2\Phi_0)(\partial \omega_{\rm q}/\partial \delta f)$ from this slope, we have to calibrate the horizontal axis. For this we need the calibration factor $\kappa \equiv \partial \delta f/\partial \delta I_\varepsilon$ [cf. (\ref{eq:transfer_function_k})], which has already been discussed above.

For the analysis of the $\Delta(\alpha)$ dependence we need a second transfer function, relating the coil current $I_{\rm coil}$ sent through the solenoid to the frustration $f_{\alpha,{\rm net}}$ of the $\alpha$-loop. With (\ref{eq:fluxoidquant3}) and (\ref{eq:flux3}) we obtain
\begin{equation}
\zeta \equiv \frac{\partial f_{\alpha,{\rm net}}}{\partial I_{\rm coil}} = \frac{A_{\mathrm{\alpha}}}{A_{\mathrm{tr}}} \;  \frac{1}{1+\beta} \, \frac{\partial f_{\rm ex}}{\partial I_{\rm coil}}
 \; .
\label{eq:transfer_function_h}
\end{equation}
With this transfer function and the expressions (\ref{eq:Ip}) and (\ref{eq:alphamax1}) for $I_{\rm p}$ and $\alpha$, respectively, we obtain
\begin{equation}
\frac{\partial \omega_{\rm q}}{\partial \delta f} = \frac{2\Phi_0 I_{\rm p}}{\hbar} = \frac{2\Phi_0 I_{\rm c}}{\hbar} \sqrt{1-\left[ 2\alpha_0 \left| \cos \left( \pi \zeta I_{\rm coil} + \pi \frac{A_\alpha}{A_{\rm tr}} \frac{\beta}{1+\beta} n \right) \right|  \right]^{-2} }
 \; .
\label{eq:transfer_omega_deltaf}
\end{equation}
Using the abbreviations $\eta = 2\Phi_0 I_c \kappa /\hbar$ and $I_n = (A_{\mathrm{\alpha}}/A_{\mathrm{tr}})(\beta/1+\beta)(n/\zeta)$ this simplifies to
\begin{equation}
\frac{\partial \omega_{\rm q}}{\partial \delta I_\varepsilon} = \eta \sqrt{1-\left[ 2\alpha_0 \left| \cos \left( \pi \zeta [I_{\rm coil} + I_n] \right) \right| \right]^{-2} }
 \; .
\label{eq:transfer_omega_delta_varepsilon}
\end{equation}
We can use this expression to fit the measured $\partial \omega_{\rm q}(I_{\rm coil})/\partial \delta I_\varepsilon$ dependence using $\eta$, $I_n$ and $\zeta$ as fitting parameters.

\begin{figure}[tb]
\center{\includegraphics[width=0.95\columnwidth]{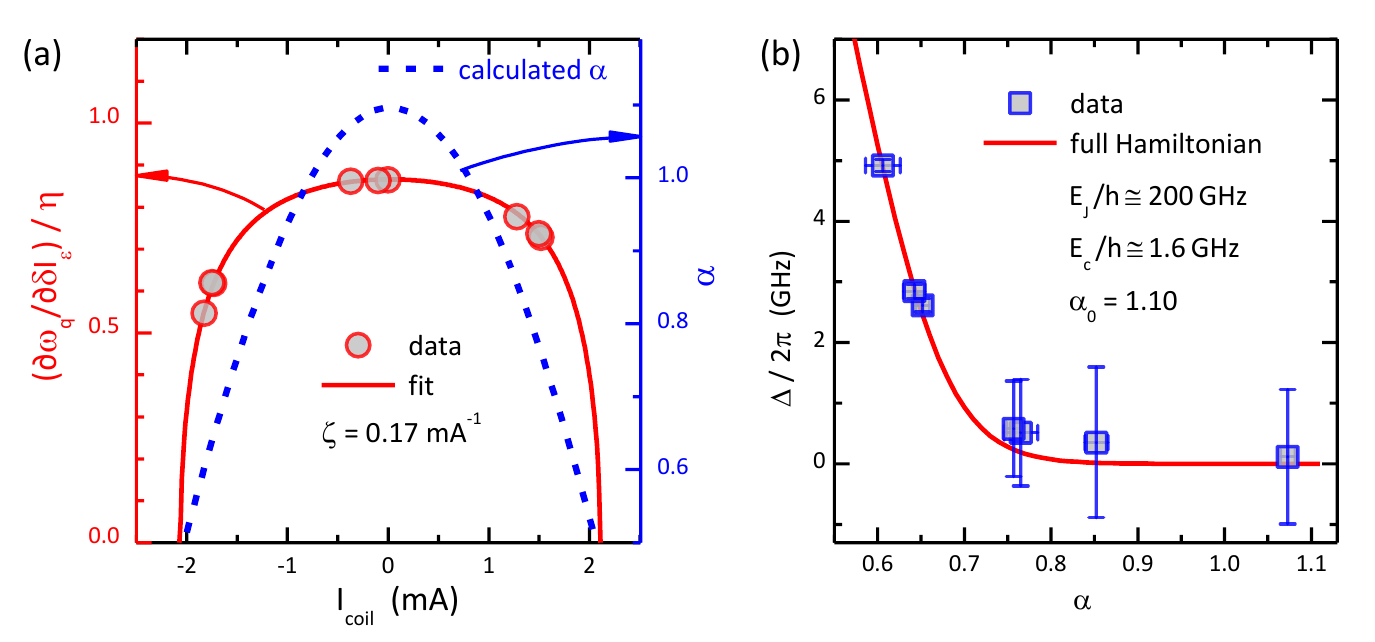}}
 \caption{
  (a) Measured $\partial \omega_{\rm q}/\partial \delta I_\varepsilon$ values plotted versus the coil current $I_{\rm coil}$ through the solenoid producing the homogeneous magnetic field for a tunable-gap gradiometric flux qubit. The solid line is a fit to the data using (\ref{eq:transfer_omega_delta_varepsilon}) yielding $\zeta$ and $I_n$. The broken line shows the calculated $\alpha (I_{\rm coil})$ dependence for these fitting parameters. (b) Minimal qubit transition frequency $\Delta /2\pi$ plotted versus $\alpha$. The solid line is obtained by numerical simulations based on the full qubit Hamiltonian using the parameters $E_{\rm J}/h = 200$\,GHz and $E_{\rm c}/h = 1.6$\,GHz.
 }
\label{Schwarz_NJP_Figure7}
\end{figure}

In figure~\ref{Schwarz_NJP_Figure7}(a) the measured $\partial \omega_{\rm q}/\partial \delta I_\varepsilon$ values are plotted versus $I_{\rm coil}$ together with a fit by (\ref{eq:transfer_omega_delta_varepsilon}). Evidently, the data points are clustered near specific $I_{\rm coil}$ values. The reason is that the homogeneous magnetic field produced by $I_{\rm coil}$ also changes the frustration of the readout SQUID and that the sensitivity of this SQUID is sufficient only in a limited range of frustration. Figure~\ref{Schwarz_NJP_Figure7}(a) shows that the expression (\ref{eq:transfer_omega_delta_varepsilon}) fits the experimental data well, yielding values for $\zeta$ and $I_n$. With these fitting parameters we can calculate $\alpha = \alpha_0 \left| \cos \left( \pi \zeta [I_{\rm coil} + I_n] \right) \right|$. The resulting curve is also shown in figure~\ref{Schwarz_NJP_Figure7}(a). We note, however, that in this case the fit parameters $I_n$ and $\zeta$ cannot be used to directly determine $\beta$ from the expression $I_n = (A_{\mathrm{\alpha}}/A_{\mathrm{tr}})(\beta/1+\beta)(n/\zeta)$, because the value of $I_n$ can be distorted by an additional background magnetic field. Therefore, we use only differences $\Delta I_n$ to determine $\beta$ [cf. (\ref{eq:beta})]. Knowing the $\alpha(I_{\rm coil})$ dependence, we can adjust $\alpha$ to any desired value by adjusting $I_{\rm coil}$ and then do spectroscopy at these values. Fitting the spectroscopy data [cf. figure~\ref{Schwarz_NJP_Figure6}(c)], we can derive the qubit gap $\Delta$ and plot it versus $\alpha$. The result is shown in figure~\ref{Schwarz_NJP_Figure7}(b) together with the dependence obtained from numerical simulations based on the full Hamiltonian. The agreement between the experimental data and the numerical simulation is best for $E_{\rm J}/h = 200$\,GHz and $E_{\rm c}/h = 1.6$\,GHz, i.e. $E_{\rm J}/E_{\rm c}= 125$. We note that the $E_{\rm J}$ value agrees well with the one estimated independently from the measured junction areas and the $J_{\rm c}$ value measured for the junctions of the readout SQUID. This clearly shows the consistency of the data analysis and demonstrates the good control on the junction parameters fabricated on the same chip. Knowing the $\Delta (\alpha)$ and $\alpha(I_{\rm coil})$ dependencies we can adjust the qubit gap \textit{in situ} by $I_{\rm coil}$, while operating the qubit at the symmetry point with optimal coherence properties. This is a key prerequisite for many applications of flux qubits.

\begin{figure}[tb]
\center{\includegraphics[width=0.95\columnwidth]{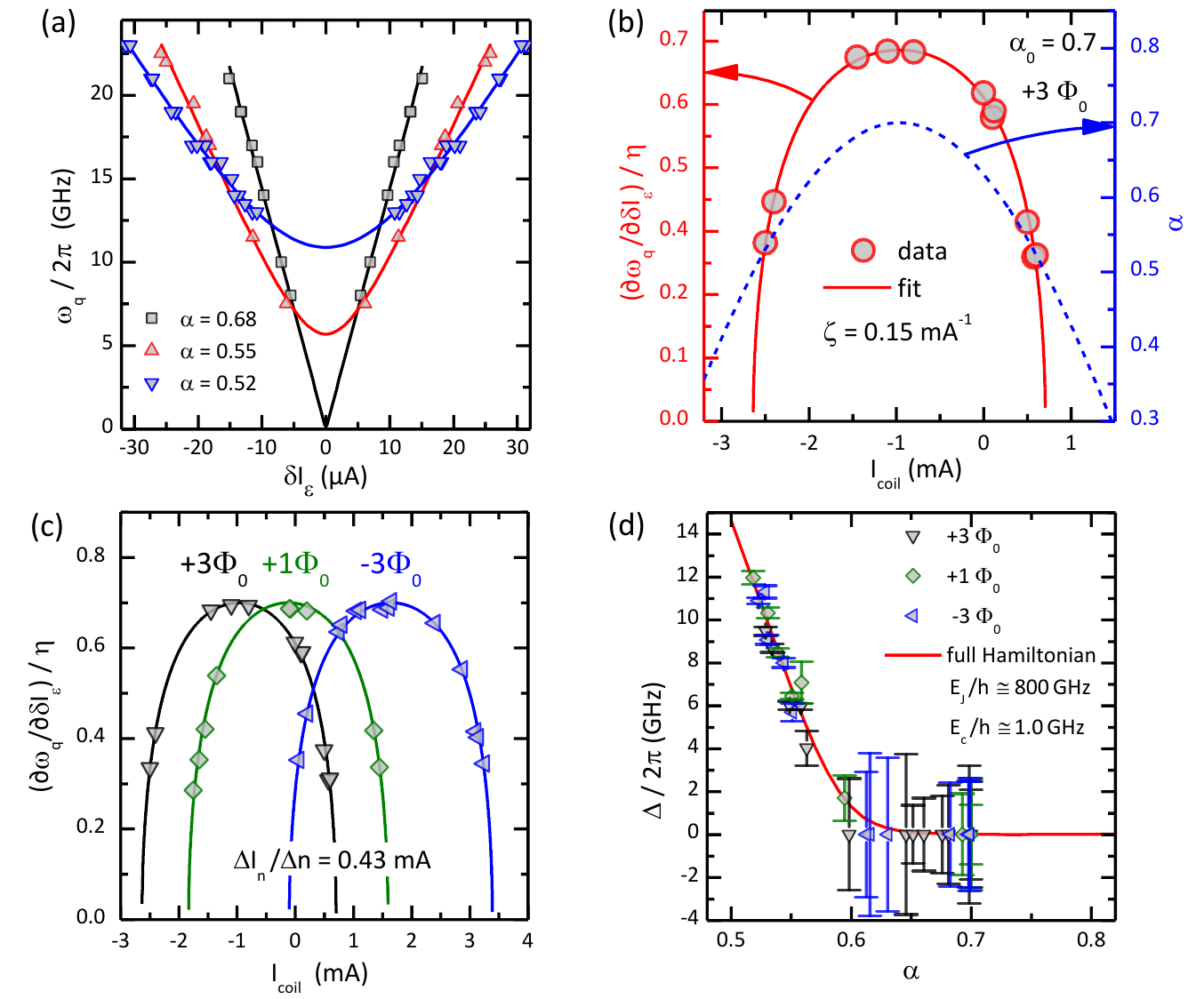}}
 \caption{
  (a) Transition frequency $\omega_{\rm q}/2\pi$ plotted versus $\delta I_\varepsilon = I_\varepsilon - I_\varepsilon^{\rm sym}$ for a tunable-gap gradiometric flux qubit with $\alpha_0 = 0.7$. Also shown is the result of a two-parameter fit. (b) Measured $\partial \omega_{\rm q}/\partial \delta I_\varepsilon$ values plotted versus the coil current $I_{\rm coil}$ through the solenoid producing the homogeneous magnetic field for $n\,=\,+3$ trapped flux quanta. The solid line is a fit to the data by (\ref{eq:transfer_omega_delta_varepsilon}) yielding $\zeta$ and $I_n$. The broken line shows the calculated $\alpha (I_{\rm coil})$ dependence for these fitting parameters. (c) Measured values as in (b) for three different values of the number of trapped flux quanta, $n=-3,+1,+3$, fitted with consistent parameters. From the horizontal displacement of the different curves we obtain $\Delta I_n/\Delta n=0.43\,$mA. (d) Minimal qubit transition frequency $\Delta /2\pi$ plotted versus $\alpha$ for three different values of trapped flux quanta. The solid line is a fit of the data based on the full qubit Hamiltonian with the fitting parameters $E_{\rm J}/h = 800$\,GHz and $E_{\rm c}/h = 1.0$\,GHz.
 }
\label{Schwarz_NJP_Figure8}
\end{figure}

For the sample of figure~\ref{Schwarz_NJP_Figure7}, the qubit gap could be varied between values close to zero and about 5\,GHz. For comparison, the data of a sample with larger ratio $E_{\rm J}/E_{\rm c}=800$ is shown in figure~\ref{Schwarz_NJP_Figure8}. The overall behaviour of this sample is very similar but the qubit gap can be tuned to values above 10\,GHz. Moreover, we investigate the tunability of this sample for different amounts of trapped flux quanta $n$. In figure~\ref{Schwarz_NJP_Figure8}(c) we plot $\partial \omega_{\rm q}/\partial \delta I_\varepsilon$ versus the coil current $I_{\rm coil}$ through the solenoid for three different values of the trapped flux ranging from $n=-3$ to $n=+3$. Evidently, the general shape of the three curves is very similar as well as the obtained fitting parameters $\zeta$ and $\eta$. The shift along the horizontal axis is expected from (\ref{eq:transfer_omega_delta_varepsilon}) and can now be used to calculate $\beta$. Starting with the expression $I_n = (A_{\mathrm{\alpha}}/A_{\mathrm{tr}})(\beta/1+\beta)(n/\zeta)$, we only use differences $\Delta I_n = I_{n,i} - I_{n,j}$. They correspond to differences $\Delta n = n_i - n_j$ and result in $\Delta I_n = (A_{\mathrm{\alpha}}/A_{\mathrm{tr}})(\beta/1+\beta)(\Delta n/\zeta)$. For our sample, we find a mean value of $\Delta I_n/\Delta n=0.43\,$mA, finally yielding
\begin{equation}
\beta = \left( \frac{\Delta n}{\Delta I_n\zeta} \frac{A_{\mathrm{\alpha}}}{A_{\mathrm{tr}}} -1 \right)^{-1} = 0.52
 \; .
\label{eq:beta}
\end{equation}
This value is in reasonable agreement with the one derived from the $L_{\rm g}$ and $L_{\rm k}$ values which can be estimated from the qubit geometry, the cross-sectional area of the superconducting lines and the dirty limit expression of $L_{\rm k}$. We note that the result from (\ref{eq:beta}) is more precise because it is computed directly from the sample. All in all, our results show that the measured data agree well with the behaviour expected from theory. Moreover, the values of $E_{\rm J}$ and $E_{\rm c}$ obtained from fitting the data agree well with those obtained for junctions fabricated on the same chip. This demonstrates that the gap of gradiometric flux qubits can be reliably tuned over a wide range, making them attractive for a large number of applications.

\begin{figure}[tb]
\center{\includegraphics[width=0.9\columnwidth]{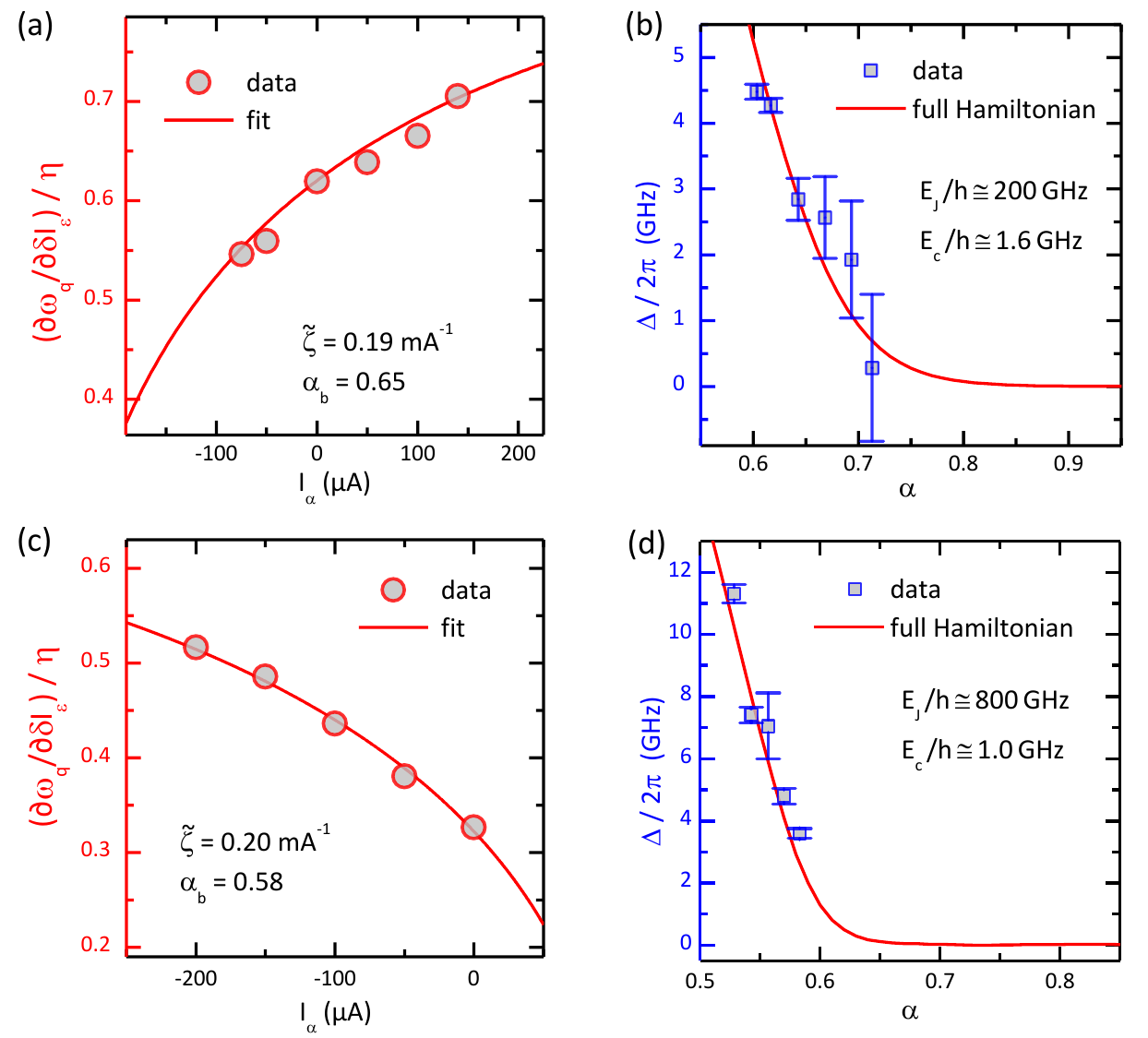}}
 \caption{
  (a) and (c) Measured $\partial \omega_{\rm q}/\partial \delta I_\varepsilon$ values plotted versus the current $I_\alpha$ through the $\alpha$-flux line for two tunable-gap gradiometric flux qubits. The solid lines are fits to the data by (\ref{eq:transfer_omega_delta_varepsilon2}) yielding  the fitting parameter $\widetilde{\zeta}$. (b) and (d) Minimal qubit transition frequency $\Delta /2\pi$ plotted versus $\alpha$ of the qubits of (a) and (c). The solid lines are obtained from numerical simulations based on the full qubit Hamiltonian using the parameters $E_{\rm J}/h$ and $E_{\rm c}/h$ as listed in the subfigures. The data in (a,b) and (c,d) are obtained for the two samples of figure~\ref{Schwarz_NJP_Figure7} and figure~\ref{Schwarz_NJP_Figure8}, respectively, however with on-chip control of $\Delta$ via the $\alpha$-flux line.
 }
\label{Schwarz_NJP_Figure9}
\end{figure}

We finally address the tuning of $\Delta$ by the on-chip $\alpha$-flux line. Since the maximum current through this line is limited by its critical current and by heating effects in contacts, only small variations of the frustration of the $\alpha$-loop are possible. Therefore, a constant applied magnetic field or a proper number of trapped flux quanta are used to pre-bias the qubit at a value $\alpha_{\rm b}$, where the slope of the $\Delta(\alpha)$ dependence is steep. Then, $I_\alpha$ is used to vary $\alpha$ around this value. In our experiments, a constant applied field is used to set $\alpha_{\rm b}$. Since the variation of the frustration of the $\alpha$-loop is generated by $I_\alpha$ instead of $I_{\rm coil}$, we have to use the modified calibration factor
\begin{equation}
\widetilde{\zeta} \equiv \frac{\partial f_{\alpha,{\rm net}}}{\partial I_\alpha}
 \; .
\label{eq:transfer_function_halpha}
\end{equation}
With this factor, we obtain
\begin{equation}
\frac{\partial \omega_{\rm q}}{\partial \delta I_\varepsilon} =  \eta \sqrt{1-\left[ 2\alpha_0 \left| \cos \left( \arccos (\alpha_{\rm b}) + \pi \widetilde{\zeta}I_\alpha \right) \right| \right]^{-2} }
 \; .
\label{eq:transfer_omega_delta_varepsilon2}
\end{equation}
We can use this expression to fit the measured $\partial \omega_{\rm q}/\partial \delta I_\varepsilon$ versus $I_\alpha$ dependence  using $\widetilde{\zeta}$ as fitting parameter. Based on these results, we can calculate $\alpha = \alpha_0 \left| \cos \left( \arccos (\alpha_{\rm b}) + \pi \widetilde{\zeta}I_\alpha \right) \right|$. Knowing $\alpha$, we can use the $\Delta$ values obtained from two-parameter fits of the spectroscopy data to get the $\Delta(\alpha)$ dependence. Experimental data for the two samples of figures~\ref{Schwarz_NJP_Figure7} and \ref{Schwarz_NJP_Figure8} are shown in figure~\ref{Schwarz_NJP_Figure9}. In figures~\ref{Schwarz_NJP_Figure9}(b) and (d), we compare the experimental $\Delta(\alpha)$ curves to numerical simulations based on the full qubit Hamiltonian with the same $E_{\rm J}$ and $E_{\rm c}$ values as obtained by tuning $\alpha$ with the coil current [cf. figure~\ref{Schwarz_NJP_Figure7}(b) and figure~\ref{Schwarz_NJP_Figure8}(d)]. The very good agreement between measurement data and calculation demonstrates again the consistency of our data analysis. All in all, our data clearly show that the qubit gap can be varied in a controlled way over a wide range by varying the frustration of the $\alpha$-loop of the gradiometric flux qubits either by an external solenoid or an on-chip control line.

\section{Summary}
\label{section_summary}

In summary, we have designed and fabricated gradiometric flux qubits with fixed and tunable gap. The characteristic parameters of the qubits have been derived from spectroscopy measurements. By trapping an odd number of flux quanta in the outer gradiometer loop during cool-down in a constant applied magnetic field, we were able to pre-bias the gradiometric flux qubits at the symmetry point. We also performed a systematic analysis of the effect of the kinetic inductance of the narrow superconducting lines forming the qubit loop. The experimental results are in good agreement with the theoretically expected behaviour. The detailed analysis of the gradiometer imbalance showed that we can fabricate gradiometric qubits with an imbalance as small as 0.2\%. This gradiometer quality is sufficient for most applications.

Since the tunability of the qubit gap of persistent current qubits is a key issue, we have performed a systematic study on the tuning of the gap of gradiometric flux qubits by external control parameters. To this end, we have replaced one of the Josephson junctions in the qubit loop by a dc SQUID. This allowed us to tune the critical current of this junction and, in turn, the qubit gap \textit{in situ} by a control flux threading the SQUID loop. The control flux was generated by three different methods: (i) an external solenoid, (ii) a persistent current frozen into the outer gradiometer loop, or (iii) a current sent through an on-chip control line. We have performed spectroscopic measurements, demonstrating a well-defined controllability of the qubit gap between values close to zero and more than 10\,GHz. Our results clearly show that it is possible to vary the qubit gap over a wide range, as it is desired for tuning in and out of resonance with superconducting quantum circuits, while operating the qubit at its symmetry point with optimal dephasing properties. Due to the steep dependence of the qubit gap on the control flux in some parameter regime, a very fast tuning of the qubit gap with small currents through on-chip control lines is feasible.

We have compared the experimental data to model calculations based on the full qubit Hamiltonian. In general, very good agreement between experiment and model calculations is achieved. Fitting the data allowed us to determine the Josephson coupling and the charging energies of the qubit junctions. The derived values agree well with those measured for single junctions or SQUIDs fabricated on the same chip. By the controlled tunability of the flux qubits a major drawback of this qubit type has been overcome. With their specific advantages such as their large anharmonicity and their potentially strong coupling to resonators, tunable-gap gradiometric flux qubits are highly attractive for the implementation of quantum information circuits or the realization of fundamental quantum experiments.

\ack
The authors gratefully acknowledge financial support from the Deutsche Forschungsgemeinschaft via the Sonderforschungsbereich~631, the German Excellence Initiative via the `Nanosystems Initiative Munich' (NIM), and the EU projects CCQED and PROMISCE.

\section*{References}

\bibliographystyle{unsrt}
%\bibliography{Schwarz_GTFQ_NJP}

\end{document}